\documentclass[12pt,final,onecolumn]{IEEEtran}
\usepackage{graphicx,psfrag,epsfig,epsf,latexsym,hhline,amsmath,amssymb}
\usepackage{hyperref}
\usepackage{graphicx}
\newtheorem{theorem}{Theorem}

\newtheorem{remark}[theorem]{Remark}
\usepackage{setspace}
\begin{document}
\title{Joint Source-Channel Coding with Correlated Interference \footnote{This paper was published in part at the 2011 IEEE International Symposium on Information Theory.}}

\author{Yu-Chih Huang and Krishna R. Narayanan\\
Department of Electrical and Computer Engineering \\
Texas A\&M University\\
{\tt\small {\{jerry.yc.huang@gmail.com, krn@ece.tamu.edu\}} }}

\maketitle

\begin{abstract}
We study the joint source-channel coding problem of transmitting a discrete-time analog source over an additive white Gaussian noise (AWGN) channel with interference known at transmitter. We consider the case when the source and the interference are correlated. We first derive an outer bound on the achievable distortion and then, we propose two joint source-channel coding schemes. The first scheme is the superposition of the uncoded signal and a digital part which is the concatenation of a Wyner-Ziv encoder and a dirty paper encoder. In the second scheme, the digital part is replaced by the hybrid digital and analog scheme proposed by Wilson \emph{et al}. When the channel signal-to-noise ratio (SNR) is perfectly known at the transmitter, both proposed schemes are shown
to provide identical performance which is substantially better than that of existing schemes. In the presence of an SNR mismatch, both proposed schemes are shown to be capable of graceful enhancement and graceful degradation. Interestingly, unlike the case when the source and interference are independent, neither of the two schemes outperforms the other universally. As an application of the proposed schemes, we provide both inner and outer bounds on the distortion region for the generalized cognitive radio channel.
\end{abstract}


%%%%%%%%%%%%%%%%%%%%%%%%%%%%%%%%%%%%%%%%%%%%%%%%%%%%%%%%%%%%%%%
\section{Introduction and Problem Statement}
In this paper, we consider transmitting a length-$n$ i.i.d. zero-mean Gaussian source $V^n=(V(1), V(2), \ldots, V(n))$ over $n$ uses of an additive white Gaussian noise (AWGN) channel with noise $Z^n\sim \mathcal{N}(0,N\cdot I)$ in the presence of Gaussian interference $S^n$ which is known at the transmitter as shown in Fig.~\ref{fig:sys_mod_1}. Throughout the paper, we only focus on the bandwidth-matched case, i.e., the signalling rate over the channel is equal to the sampling rate of the source. The transmitted signal $X^n=(X(1), X(2), \ldots, X(n))$ is subject to a power constraint
\begin{equation}
    \frac{1}{n}\sum^n_{i=1}\mathbb{E}[X(i)^2] \leq P,
\end{equation}
where $\mathbb{E[\cdot]}$ represents the expectation operation. The received signal $Y^n$ is given by
\begin{equation}\label{eqn:DPC_setup}
    Y^n = X^n + S^n + Z^n.
\end{equation}

We are interested in the expected distortion between the source and the estimate $\widehat{V}^n$ at the output of the decoder given by
\begin{equation}
    d = \mathbb{E}[d(V^n,g(f(V^n,S^n) + S^n + Z^n))],
\end{equation}
where $f$ and $g$ are a pair of source-channel coding encoder and decoder, respectively, and $d(. , .)$ is the mean squared error (MSE) distortion measure given by
\begin{equation}\label{eqn:MSE}
    d(\mathbf{v}, \hat{\mathbf{v}}) = \frac{1}{n}\sum^n_{i=1}(v(i)-\hat{v}(i))^2.
\end{equation}
Here the lower case letters represent realizations of random variables denoted by upper case letters. As in \cite{Cover}, a distortion $D$ is achievable under power constraint $P$ if for any $\varepsilon > 0$, there exists a source-channel code and a sufficiently large $n$ such that $d \leq D+\varepsilon$.

When $V$ and $S$ are uncorrelated, it is known that an optimal quantizer followed by a Costa's dirty paper coding (DPC) \cite{DPC} is optimal and the corresponding joint source-channel coding problem is fully discussed in \cite{Makesh}. However, different from the typical writing on dirty paper problem, in this paper, we consider the case where the source and the interference are correlated with a covariance matrix given by
\begin{equation}\label{eqn:CovSV}
    \Lambda_{VS} = \left(
                 \begin{array}{cc}
                   \sigma_V^2 & \rho\sigma_V\sigma_S \\
                   \rho\sigma_V\sigma_S & \sigma_S^2 \\
                 \end{array}
               \right).
\end{equation}

Under this assumption, separate source and channel coding using DPC naively may not be a good candidate for encoding $V^n$ in general. It is due to the fact that in Costa's DPC scheme, the transmitted signal is designed to be orthogonal to the interference and, hence, the DPC scheme cannot exploit the correlation between the source and the interference. Also, the purely uncoded scheme fails to avoid the interference and is suboptimal in general. In this paper, we first derive an outer bound on the achievable distortion region and then, we propose two joint source-channel coding schemes which exploit the correlation between $V^n$ and $S^n$, thereby outperforming the naive DPC scheme. The first scheme is a superposition of the uncoded scheme and a digital part formed by a Wyner-Ziv coding \cite{Wyner-Ziv} followed by a DPC, which we refer to as a digital DPC based scheme (or just the digital DPC scheme). The second scheme is obtained by replacing the digital part by a hybrid digital and analog (HDA) scheme given in \cite{Makesh} that has been shown to provide graceful improvement when the actual SNR (SNR$_a$) is better than the design SNR (SNR$_d$). We then analyze the performance of these two proposed schemes when there is an SNR mismatch. It is shown that both the HDA scheme and the digital DPC scheme benefit from a higher channel SNR and provide graceful enhancement; however, interestingly, for this case neither of schemes dominate the other universally and which one performs better depends on the designed SNR. When $\rho$ is small, the HDA scheme outperforms the digital DPC scheme and when $\rho$ is large, the digital DPC scheme outperforms the HDA scheme. When the channel deteriorates, both the proposed schemes perform identically and are able to provide graceful degradation.

One interesting application of this problem is to derive an achievable distortion region for the generalized cognitive radio channel with correlated sources. This channel can be modeled as a typical two-user interference channel except that one of them knows exactly what the other plans to transmit. Moreover, two users' sources are assumed to be correlated. One can regard the informed user's channel as the setup we consider here and then directly apply the schemes we propose as the coding scheme for the informed user. For the generalized cognitive radio channel with correlated sources, we provide inner and outer bounds on the distortion region where the inner bound largely relies on the coding schemes proposed in this paper.

The rest of the paper is organized as follows. In Section \ref{sec:JSCC}, we present some prior work which is closely related to ours. The outer bound is given in Section \ref{sec:JSCC_ob} and two proposed schemes are given in Section \ref{sec:proposed_scheme}. In Section \ref{sec:SNR_mismatch}, we analyze the performance of the proposed schemes under SNR mismatch. These proposed schemes are then extended to the generalized cognitive radio channel in Section \ref{sec:JSCC_cograd}. Some conclusions are given in Section \ref{sec:conclusions}.

%%%%%%%%%%%%%%%%%%%%%%%%%%%%%%%%%%%%%%%%%%%%%%%%%%%%%%%%%%%%%%%
\section{Related Work on JSCC with Interference Known at Transmitter} \label{sec:JSCC}
In \cite{Sutivong}, Sutivong \emph{et al.} consider the problem of sending a digital source in the presence of interference (or, channel state) which is known at the transmitter and is assumed to be independent of the source. The optimal tradeoff between the achievable rate for transmitting the digital source and the distortion in estimating the interference is then studied. A coding scheme that is able to achieve the optimal tradeoff is also provided in \cite{Sutivong}. This coding scheme uses a portion of the power to amplify the interference and uses the remaining power to transmit the digital source via DPC. This coding scheme can be extended to the problem we consider as follows. Since the source and the interference are jointly Gaussian, we can first rewrite the source as $V = \rho \frac{\sigma_V}{\sigma_S}S + N'_{\rho}$ with $S$ and the innovation $N'_{\rho}$ being independent of each other. Now if one quantizes $N'_{\rho}$ into digital data, the setup becomes the one considered by Sutivong \emph{et al.} and their proposed scheme can be applied directly. For any power allocation between the analog part and digital part, using this scheme to operate on the boundary of the optimal tradeoff, the optimal distortion in estimating $\rho\frac{\sigma_V}{\sigma_S}S$ and that in estimating $N'_{\rho}$ is achieved. The distortion in estimating $V$ for this power allocation strategy is the sum of the above two distortions. One can then optimize the power allocation strategy to get the minimum distortion for this coding scheme. It is worth pointing out that this coding scheme is in general suboptimal for our problem although it achieves the optimal tradeoff between estimating $S$ and $N'_{\rho}$ individually. This is because, our interest is in estimating $V$ directly and it is importantly to carefully take advantage of the correlation in the estimation error in estimating $S$ and $N_{\rho}'$. The coding scheme in \cite{Sutivong} is not naturally suited to take advantage of this correlation. One numerical example is shown in Fig.~\ref{fig:Sutivong} where we can see that the union of the uncoded scheme and the naive DPC scheme outperforms the extension of Sutivong \emph{et al.}'s scheme.

In \cite{LapTin}, Lapidoth \emph{et al.} consider the $2 \times 1$ multiple access channel in which two transmitters wish to communicate their sources, which are drawn from a bi-variate Gaussian distribution, to a receiver which is interested in reconstructing both sources. There are some similarities between the proposed work and the work in \cite{LapTin} if we regard one of the users', say the user 2's, signal as interference. However, an important difference is that in \cite{LapTin}, the transmitters are not allowed to cooperate with each other, i.e., for the transmitter 1, the interference (user 2's signal) is not known. Moreover, this interference now depends on the signalling scheme adopted at user 2 and may not be correlated to the source anymore.

In \cite{BrossLapTin08_BC}-\cite{TianDigSha}, transmitting a bi-variate Gaussian source over a $1\times 2$ Gaussian Broadcast Channel is considered. In their setup, the source consists of two components $V^n_1$ and $V^n_2$ which are memoryless and stationary bi-variate Gaussian random variables and each receiver is only interested in one part of the sources. In \cite{TianDigSha}, Tian \emph{et al.} propose a HDA scheme that achieves the outer bound given in \cite{BrossLapTin08_BC} and therefore leads to a complete characterization of the distortion region. This problem is similar to ours if we only focus on one receiver, say the first receiver. However, a crucial difference is that the interference now is a function of $V^n_2$ which depends on the broadcast encoding scheme and may not be correlated to $V^n_1$. The joint source-channel coding problem for broadcasting a single memoryless Gaussian source under bandwidth mismatch is considered in \cite{Mittal02}-\cite{Tian11}. However, different from its bandwidth matched counterpart \cite{Goblick}, only approximation characterizations of the achievable distortion region are available for this problem. Broadcasting a colored Gaussian source over a colored Gaussian broadcast channel to a digital receiver and a analogy receiver is considered in \cite{Puri08} where Prabhakaran \emph{et al.} propose a HDA scheme that achieves the entire distortion region for the problem they consider.

Joint source-channel coding for point to point communications over Gaussian channels has also been widely discussed. See e.g. \cite{Makesh},\cite{ShamaiVerduZamir}-\cite{BrossLapTin06}. However, they either don't consider interference (\cite{ShamaiVerduZamir}-\cite{BrossLapTin06}) or assume independence of source and interference (\cite{Makesh}). In \cite{Makesh}, Wilson \emph{et al.} proposed a HDA coding scheme for the typical writing on dirty paper problem in which the source is independent of the interference. This HDA scheme was originally proposed to perform well in the case of a SNR mismatch. In \cite{Makesh}, the authors showed that their HDA scheme not only achieves the optimal distortion in the absence of SNR mismatch but also provides gracefully degradation in the presence of SNR mismatch. In the following sections, we will discuss this scheme in detail and then propose a coding scheme based on this one.
%In this section , we will adopt this encoding scheme with different coefficients to exploit the correlation between $V$ and $S$.

%%%%%%%%%%%%%%%%%%%%%%%%%%%%%%%%%%%%%%%%%%%%%%%%%%%%%%%%%%%%
\section{Outer Bounds}\label{sec:JSCC_ob}
\subsection{Outer Bound 1}
For comparison, we first present a genie-aided outer bound. This outer bound is derived in a similar way to the one in \cite{SouVis} in which we assume that $S^n$ is revealed to the decoder by a genie. Thus, we have
\begin{align}\label{eqn:ob_derive}
    \frac{n}{2}\log \frac{\sigma_V^2(1-\rho^2)}{D_{ob}} &\overset{(a)}{\leq} I(V^n;\widehat{V}^n|S^n) \nonumber \\
    &\overset{(b)}{\leq} I(V^n;Y^n|S^n) \nonumber \\
    &= h(Y^n|S^n) - h(Y^n|S^n,V^n) \nonumber \\
    &= h(X^n+Z^n|S^n) - h(Z^n) \nonumber \\
    &\overset{(c)}{\leq} h(X^n+Z^n) - h(Z^n) \nonumber \\
    &\overset{(d)}{\leq} \frac{n}{2}\log\left(1 + \frac{P}{N}\right),
\end{align}
where (a) follows from the rate-distortion theorem \cite{Cover}, (b) is from the data processing inequality, (c) is due from that conditioning reduces differential entropy and (d) comes from the fact that Gaussian density maximizes the differential entropy and all random variables involved are i.i.d. Therefore, we have the outer bound as
\begin{equation}\label{eqn:ob_result1}
D_{ob,1} = \frac{\sigma_V^2(1-\rho^2)}{1 + P/N}.
\end{equation}

Note that this outer bound in general may not be tight for our setup since in the presence of correlation, giving $S^n$ to the decoder also offers a correlated version of the source that we wish to estimate. For example, in the case of $\rho = 1$, giving $S^n$ to the decoder implies that the outer bound is $D_{ob} = 0$ no matter what the received signal $Y^n$ was. On the other hand, if $\rho = 0$, the setup reduces to the one with uncorrelated interference and we know that this outer bound is tight. Now, we present another outer bound that improves this outer bound for some values of $\rho$.

\subsection{Outer Bound 2}
Since $S(i)$ and $V(i)$ are jointly Gaussian distributed with covariance matrix given in \eqref{eqn:CovSV}, we can write
\begin{equation}
    S(i) = \rho\frac{\sigma_S}{\sigma_V}V(i) + N_{\rho}(i),
\end{equation}
where $N_{\rho}(i)\sim\mathcal{N}\left( 0, (1-\rho^2)\sigma_S^2 \right)$ representing the innovation and is independent to $V(i)$. Now, suppose a genie reveals only the $n$-letter collection of innovation $N^n_{\rho}$ to the decoder, we have
\begin{align}\label{eqn:ob_derive2}
    \frac{n}{2}\log &\frac{\sigma_V^2}{D_{ob,2}} =\frac{n}{2}\log \frac{\mathbf{var}(V|N_{\rho})}{D_{ob}} \nonumber \\
    &\overset{(a)}{\leq} I(V^n;\widehat{V}^n|N^n_{\rho}) \nonumber \\
    &\overset{(b)}{\leq} I(V^n;Y^n|N^n_{\rho}) \nonumber \\
    &=h(Y^n|N^n_{\rho}) - h(Y^n|N^n_{\rho},V^n) \nonumber \\
    &= h(X^n+\rho\frac{\sigma_S}{\sigma_V}V^n+Z^n|N^n_{\rho}) - h(Z^n) \nonumber \\
    &\overset{(c)}{\leq} h(X^n+\rho\frac{\sigma_S}{\sigma_V}V^n+Z^n) - h(Z^n) \nonumber \\
    &\overset{(d)}{\leq} \frac{n}{2}\log\left(\frac{\mathbf{var}\left(X+\rho\frac{\sigma_S}{\sigma_V}V + Z \right)}{N}\right) \nonumber \\
    &\overset{(e)}{\leq} \frac{n}{2}\log\left( 1 + \frac{(\sqrt{P}+\rho\sqrt{\sigma_S^2})^2}{N} \right),
\end{align}
where (a)-(d) follow from the same reasons with those in the previous outer bound and (e) is due from the Cauchy-Schwartz inequality that states that the maximum occurs when $X$ and $V$ are collinear. Thus, we have
\begin{equation}\label{eqn:ob_result2}
D_{ob,2} = \frac{\sigma_V^2}{1 + (\sqrt{P}+\rho\sqrt{\sigma_S^2})^2/N}.
\end{equation}
Note that although the encoder knows the interference $S^n$ exactly instead of just $N^n_{\rho}$, the inequality in step (a) does not decrease the knowledge about $S^n$ at the transmitter since $S^n$ is a deterministic function of $V^n$ and $N^n_{\rho}$.

\begin{remark}
    If $\rho = 0$, this outer bound reduces to the previous one and is tight. If $\rho = 1$, the genie actually reveals nothing to the decoder and the setup reduces to the one considered in \cite{Sutivong}, i.e., the encoder is interested in revealing the interference to the decoder. For this case, we know that this outer bound is tight. However, this outer bound is in general optimistic except for two extremes. It is due to the fact that in derivations, we assume that we can simultaneously ignore the $N^n_{\rho}$ and use all the power to take advantage of the coherent part. Despite this, the outer bound still provides an insight that in order to build a good coding scheme that one should try to use a portion of power to make use of the correlation and then use the remaining power to avoid $N^n_{\rho}$. Further, it is natural to combine these two outer bounds as $D_{ob} = \max\{D_{ob,1},D_{ob,2}\}$.
\end{remark}

From now on, since the channel we consider is discrete memoryless and all the random variables we consider are i.i.d. in time, i.e. $V(i)$ is independent of $V(j)$ for $i \neq j$, we will drop the index $i$ for the sake of convenience.

%%%%%%%%%%%%%%%%%%%%%%%%%%%%%%%%%%%%%%%%%%%%%%%%%%%%%%%%%%%%%%%%%%%%%%%%%%%%%%%%
\section{Proposed Schemes}\label{sec:proposed_scheme}

\subsection{Digital DPC Based Scheme}\label{subsec:DigitalDPC}
We now propose a digital DPC scheme which retains the advantages of the above two schemes. This scheme can be regarded as an extended version of the coding scheme in \cite{Puri} to the setup we consider. As shown in Fig.~\ref{fig:sep_scheme}, the transmitted signal of this scheme is the superposition of the analog part $X_a$ with power $P_a$ and the digital part $X_d$ with power $P-P_a$. The motivation here is to allocate some power for the analog part to make use of the interference which is somewhat coherent to the source for large $\rho$'s and to assign more power to the digital part to avoid the interference when $\rho$ is small. The analog part is the scaled version of linear combination of source and interference as
\begin{equation}\label{eqn:analog}
    X_a = \sqrt{a}\left(\gamma V + (1-\gamma) S\right),
\end{equation}
where $P_a\in[0,P]$, $a =P_a/\sigma_a^2$, $\gamma\in[0,1]$ and
\begin{equation}
    \sigma_a^2 = \gamma^2\sigma_V^2 + (1-\gamma)^2\sigma_S^2+2\gamma(1-\gamma)\rho\sigma_V\sigma_S.
\end{equation}

The received signal is given by
\begin{align}
    Y &= X_d + X_a + S + Z \nonumber \\
    &= X_d + \sqrt{a}\gamma V + \left( 1 + \sqrt{a}(1-\gamma) \right) S + Z \nonumber \\
    &= X_d + S' + Z,
\end{align}
where $X_d$ is chosen to be orthogonal to $S$ and $V$ and $S'=\sqrt{a}\gamma V + \left( 1 + \sqrt{a}(1-\gamma) \right) S$ is the effective interference. The receiver first makes an estimate from $Y$ only as $V' = \beta Y$ with
\begin{equation}\label{eqn:beta}
    \beta = \frac{\mathbb{E}[VY]}{\mathbb{E}[Y^2]}=\frac{\sqrt{a}(\gamma\sigma_V^2 + (1-\gamma)\rho\sigma_V\sigma_S) + \rho\sigma_V\sigma_S}{P + N + \sigma_S^2 + + 2\sqrt{a}\left((1-\gamma)\sigma_S^2 + \gamma\rho\sigma_V\sigma_S\right)}.
\end{equation}
The corresponding MSE is
\begin{align}\label{eqn:Ds_sep}
    D^* &= \sigma_V^2 - \beta\mathbb{E}[VY] \nonumber \\
    &=\sigma_V^2\left[ 1 - \beta\left( \sqrt{a}(\gamma + (1-\gamma)\rho\frac{\sigma_S}{\sigma_V} )+\rho\frac{\sigma_S}{\sigma_V}  \right)  \right].
\end{align}
Thus, we can write $V = V' + W$ with $W\sim \mathcal{N}(0,D^*)$.

We now refine the estimate through the digital part, which is the concatenation of a Wyner-Ziv coding and a DPC. Since the DPC achieves the rate equal to that when there is no interference at all, the encoder can use the remaining power $P-P_a$ to reliably transmit the refining bits $T$ with a rate arbitrarily close to
\begin{equation}\label{eqn:channel_rate}
    R = \frac{1}{2}\log\left( 1+\frac{P-P_a}{N} \right).
\end{equation}
The resulting distortion after refinement is then given as
\begin{equation}\label{eqn:D_sep}
    D_{sep} = \underset{\gamma,~P_a}{\inf} \frac{D^*}{1+\frac{P-P_a}{N}}.
\end{equation}
In Appendix \ref{app:DigitalWZ}, for self-containedness, we briefly summarize the digital Wyner-Ziv scheme to illustrate how to achieve the above distortion.

It is worth noting that setting $\gamma=1$ gives us the lowest distortion always. i.e., super-imposing $S$ onto the transmitted signal is completely unnecessary. However, it is in general not true for the cognitive radio setup. We will discuss this in detail in section \ref{sec:JSCC_cograd}.

\begin{remark}
    Different from the setup considered in \cite{Puri} that the optimal distortion can be achieved by any power allocation between coded and uncoded transmissions, in our setup the optimal distortion is in general achieved by a particular power allocation which is a function of $\rho$.
\end{remark}

\subsection{HDA Scheme}\label{subsec:HDA}
Now, let us focus on the HDA scheme obtained by replacing the digital part in Fig.~\ref{fig:sep_scheme} by the HDA scheme given in \cite{Makesh}. The analog signal remains the same as in \eqref{eqn:analog} and the HDA output is referred to as $X_h$. Therefore, we have $Y = X_h + S' + Z$. Again, the HDA scheme regards $S'$ as interference and $V'$ described previously as side-information. The encoding and decoding procedures are similar to that in \cite{Makesh} but the coefficients need to be re-derived to fit our setup (the reader is referred to \cite{Makesh} for details).

Let the auxiliary random variable $U$ be
\begin{equation}\label{eqn:auxiliaryU}
    U = X_h + \alpha S' + \kappa V,
\end{equation}
where $X_h\sim \mathcal{N}(0,P_h)$ independent to $S'$ and $V$ and $P_h = P - P_a$. The covariance matrix of $S'$ and $V$ can be computed by \eqref{eqn:CovSV}.

\emph{Codebook Generation}: Generate a random i.i.d. codebook $\mathcal{U}$ with $2^{nR_1}$ codewords, reveal the codebook to both transmitter and
receiver.

\emph{Encoding}: Given realizations $\mathbf{s}'$ and $\mathbf{v}$, find a $\mathbf{u}\in\mathcal{U}$ such that ($\mathbf{s}',\mathbf{v},\mathbf{u}$) is jointly typical. If such a $\mathbf{u}$ can be found, transmit $\mathbf{x}_h = \mathbf{u} - \alpha \mathbf{s}' - \kappa \mathbf{v}$. Otherwise, an encoding failure is declared.

\emph{Decoding}: The decoder looks for a $\hat{\mathbf{u}}$ such that $(\mathbf{y},\mathbf{v}',\hat{\mathbf{u}})$ is jointly typical. A decoding failure is declared if none or more than one such $\hat{\mathbf{u}}$ are found. It is shown in \cite{Makesh} that if $n\rightarrow \infty$ and the condition given in \eqref{eqn:R1con2} is satisfied, the probability of $\hat{\mathbf{u}}\neq \mathbf{u}\rightarrow 0$.

\emph{Estimation}: After decoding $\mathbf{u}$, the receiver forms a linear MMSE estimate of $\mathbf{v}$ from $\mathbf{y}$ and $\mathbf{u}$. The distortion is then obtained as
\begin{equation}\label{eqn:D_hda}
    D_{hda} = \underset{\gamma,~P_a}{\inf}\left[ \sigma_V^2 - \Gamma^T\Lambda_{UY}^{-1}\Gamma\right],
\end{equation}
where $\Lambda_{UY}$ is the covariance matrix of $U$ and $Y$, and $\Gamma = \left[\mathbb{E}[VU] , \mathbb{E}[VY]\right]^T$.

In the encoding step, to make sure the probability of encoding failure vanishes with increasing $n$, we require
\begin{align}\label{eqn:R1con1}
    R_1 &> I(U;S',V) \nonumber \\
    &=h(U) - h(X_h + \alpha S' + \kappa V|S',V) \nonumber \\
    &\overset{(a)}{=} h(U) - h(X_h) \nonumber \\
    &= \frac{1}{2}\log\frac{\mathbb{E}[U^2]}{P_h},
\end{align}
where (a) follows because $X_h$ is independent of $S'$ and $V$.

Further, to guarantee the decodability of $U$ in the decoding step, one requires
\begin{align}\label{eqn:R1con2}
    R_1 &\overset{(a)}{<} I(U;Y,V') \nonumber \\
    &= h(U) - h(U|Y,V') \nonumber \\
    &= h(U) - h(U-\alpha Y - \kappa V'|Y,V') \nonumber \\
    &\overset{(b)}{=} h(U) - h(\kappa W + (1-\alpha)X_h - \alpha Z|Y),
\end{align}
where (a) follows from the error analysis of $\mathcal{E}_3$ in Section III of \cite{LimMineroKim} and (b) is due to the fact that $V'=\beta Y$. By choosing
\begin{equation}\label{eqn:alp}
    \alpha = \frac{P_h}{P_h + N},~~\kappa^2 = \frac{P_h^2}{(P_h + N)D^*},
\end{equation}
one can verify that \eqref{eqn:R1con1} and \eqref{eqn:R1con2} are satisfied. Note that in \eqref{eqn:R1con1} what we really need is $R_1 \geq I(U;S',V) + \varepsilon$ and in \eqref{eqn:R1con2} it is $R_1 \leq I(U;Y,V') - \delta$. However, since $\varepsilon$ and $\delta$ can be made arbitrarily small, these are omitted for the sake of convenience and to maintain clarity.

\begin{remark}
It is shown in Appendix~\ref{app:same} that the distortions in \eqref{eqn:D_sep} and \eqref{eqn:D_hda} are exactly the same. However, as we will see in the next section, two schemes perform differently when $\text{SNR}_a > \text{SNR}_d$.
\end{remark}

\subsection{Numerical Results}
In Fig.~\ref{fig:SNR_vs_D}, we plot the distortion (in $-10\log_{10}(D)$) for coding schemes and outer bounds described above as a function of SNR. In this figure, we set $\sigma_V^2 = \sigma_S^2 = 1$ and $\rho = 0.3$. Note that for this choice of $\sigma_V^2$, what we plot is actually the signal-to-distortion ratio. As expected, the two proposed schemes have exactly the same performance. Moreover, for this case, these two schemes not only outperform others but also approach the outer bound (maximum of two) very well.

We then fix the SNR and plot the distortion as a function of $\rho$ in Fig.~\ref{fig:rho_vs_D}. The parameters are set to be $\sigma_V^2 = \sigma_S^2 = 1$, $P = 10$, and $N = 1$.  It can be seen that both the proposed schemes perform exactly the same and that the achievable distortion region with the proposed scheme is larger than what is achievable with a separation based scheme using DPC and a uncoded scheme. Further, although the proposed schemes perform close to the outer bound over a wide range of $\rho$s, the outer bound and the inner bound do not coincide however, leaving room for improvement either of the outer bound or the schemes.

%%%%%%%%%%%%%%%%%%%%%%%%%%%%%%%%%%%%%%%%%%%%%%%%%%%%%%%%%%%%%%%
\section{Performance Analysis in the Presence of SNR Mismatch}\label{sec:SNR_mismatch}
In this section, we study the distortions for the proposed schemes in the presence of SNR mismatch i.e., we consider the scenario where instead of knowing the exact channel SNR, the transmitter only knows a lower bound on the channel SNR. Specifically, we assume that the actual channel noise to be $Z_a \sim \mathcal{N}(0 , N_a)$ but the transmitter only knows that $N_a\leq N$ so that it designs the coefficients assuming the noise variance is $N$. In what follows, we analyze the performance for both proposed schemes under the above assumption.

\subsection{Digital DPC Based Scheme}
Since the transmitter designs its coefficients for $N$, it aims to achieve the distortion $D_{sep}$ given in \eqref{eqn:D_sep}. It first quantizes the source to $T$ by a Wyner-Ziv coding with side-information $D^*$ given in \eqref{eqn:Ds_sep} and then encodes the quantization output by a DPC with a rate
\begin{equation}
    R = \frac{1}{2}\log \left( 1 + \frac{P-\tilde{P}_a}{N} \right),
\end{equation}
where $\tilde{P}_a$ is the power allotted to $X_a$ such that the distortion in the absence of SNR mismatch is minimized. i.e.,
\begin{equation}
    \tilde{P}_a =  \underset{P_a}{\arg\inf}  \frac{D^*}{1+\frac{P-P_a}{N}}.
\end{equation}

At receiver, since $N_a\leq N$, the DPC decoder can correctly decode $T$ with high probability. Moreover, the receiver forms the MMSE estimate of $V$ from $Y$ as $V'_a = \beta_a Y$ with $\beta_a$ and the corresponding MSE $D^*_a$ derived by substituting $N_a$ for $N$ in \eqref{eqn:beta} and \eqref{eqn:Ds_sep}, respectively.
%\begin{equation}
%    \beta_a = \frac{\sqrt{a}(\gamma\sigma_V^2 + (1-\gamma)\rho\sigma_V\sigma_S) + \rho\sigma_V\sigma_S}{P + N_a + \sigma_S^2 + 2\sqrt{a}\left((1-\gamma)\sigma_S^2 + \gamma\rho\sigma_V\sigma_S\right)},
%\end{equation}
%\begin{equation}
%    D^*_a = \sigma_V^2\left[ 1 - \beta_a\left( \sqrt{a}(\gamma + (1-\gamma)\rho\frac{\sigma_S}{\sigma_V} )+\rho\frac{\sigma_S}{\sigma_V}  \right)  \right].
%\end{equation}
After that, the problem reduces to the Wyner-Ziv problem with mismatched side-information. In Appendix~\ref{app:WZmismatch}, we show that for this problem, one can achieve
\begin{equation}\label{eqn:D_sep_mis}
    D_{sep,mis} = \frac{D^*D^*_a }{D^*D^*_a + (D^*-D^*_a)D_{sep}} D_{sep}.
\end{equation}

Unlike the typical separation-based scheme that we have seen in \cite{Makesh}, the proposed digital DPC scheme (whose digital part can be regarded as a separation-based scheme) can still take advantage of better channels through mismatched side-information.

\subsection{HDA Scheme}
Different from the digital DPC scheme, in the presence of SNR mismatch, the performance analysis of the HDA scheme cannot be converted to the Wyner-Ziv problem with mismatched side-information. It is because that in the HDA scheme, we jointly form an estimate of $V$ from $U$ and $Y$. Fortunately, as shown in \cite{Makesh}, the HDA scheme is capable of making use of an SNR mismatch.

%Although it is shown in Appendix \ref{app:WZmismatch} that the performance of the HDA scheme is exactly the same as the digital Wyner-Ziv scheme under side-information mismatch, this problem with HDA scheme cannot be reduced to the Wyner-Ziv problem with mismatch side-information as we did for the superposition-based scheme. It is because that the HDA scheme still makes an estimate of $V$ from $U$ which is a function of $S$. Fortunately, as shown in \cite{Makesh}, the HDA scheme is capable of making use of SNR mismatch.

Similar to the digital DPC scheme, we design the coefficients for noise variance $N$. The HDA scheme regards $D^*$ as side-information and $S'$ as interference. It generates the auxiliary random variable $U$ given by \eqref{eqn:auxiliaryU} with coefficients described by \eqref{eqn:alp}. Since $N_a\leq N$, the receiver can correctly decode $U$ with high probability. The receiver then forms the MMSE as described in \eqref{eqn:D_hda}. Note that $\mathbb{E}[Y^2]$ in $\Lambda_{UY}$ should be modified appropriately to address the fact that the actual noise variance is $N_a$ in this case.

\begin{remark}\label{rmk:Sutivong}
    In \cite{Makesh}, the authors compare the distortions of the digital scheme and the HDA scheme in estimating the source $V$ and the interference $S$ as we move away from the designed SNR. One important observation is that the HDA scheme outperforms the separation-based scheme in estimating the source; however, the separation-based scheme is better than the HDA scheme if one is interested in estimating the interference. Here, since the \emph{effective} interference $S'$ includes the uncoded signal $\sqrt{a}V$ in part and the source is correlated to the interference, estimating the source $V$ is equivalent to estimating a part of $S'$. Thus, one can expect that if $P_a$ and $\rho$ are large enough, the digital DPC scheme may outperform the HDA scheme in the presence of SNR mismatch. One the other hand, if $P_a$ and $\rho$ are relatively small, one can expect the reverse.

    %the benefit coming from using the HDA scheme to estimate the source may be less than that from adopting the superposition-based scheme to estimate a part of $S'$. Consequently, for a sufficiently large $P_a$ and $\rho$, the superposition-based scheme may be better than the HDA scheme in the presence of SNR mismatch.
\end{remark}

\begin{remark}
    Note that we have only discussed the case when the actual channel turns out to be better than that expected by the transmitter. On the other hand, when the channel deteriorates, the digital DPC scheme and the HDA scheme are not able to decode the digital part and the HDA part, respectively. For the digital DPC scheme, this is due to the fact that a capacity-approaching code is used so that the decoding will fail if the channel is no longer being able to support this rate. For the HDA scheme, this inability to decode $U$ is because the constraint \eqref{eqn:R1con2} is no longer satisfied if the channel is worse than that expected. However, both schemes can still form the MMSE estimate of the source from the received signal $Y$. Therefore, for a same choice of $P_a$, the resulting distortion of two proposed schemes would be the same and is equal to $D_a^*$. This implies that both the proposed schemes are able to provide graceful degradation when channel deteriorates.
\end{remark}

\subsection{Numerical Results}
Now, we compare the performance of the above two schemes and the scheme that knows the actual SNR. The parameters are set to be $\sigma_V^2 = \sigma_S^2 = 1$. We plot $-10\log_{10}(D)$ as we move away from the designed SNR for both small ($\rho = 0.1$) and large ($\rho = 0.5$) correlations. Two examples for designed SNR = 0 dB and 10 dB are given in Fig.~\ref{fig:JSCC_mismatch_small} and Fig.~\ref{fig:JSCC_mismatch_large}, respectively.

In Fig.~\ref{fig:JSCC_mismatch_small}, we consider the case that the designed SNR is 0 dB which is relatively small compared to the variance of interference. For this case, we can see that which scheme performs better in the presence of SNR mismatch really depends on $\rho$. It can be explained by the observations made in \emph{Remark}~\ref{rmk:Sutivong} and the power allocation strategy. For this case the optimal power allocation $\tilde{P}_a$ is proportional to $\rho$. For $\rho = 0.1$ case, since the correlation is small and the assigned $\tilde{P}_a$ is also small, the HDA scheme is better than the digital DPC scheme. On the other hand, for $\rho = 0.5$ case, we allot a relatively large power to $\tilde{P}_a$ so that one may get a better estimate if we try to use the digital DPC scheme to estimate a part of $S'$. This property is further discussed in the Appendix \ref{app:DiscussMismatch}.

In Fig.~\ref{fig:JSCC_mismatch_large}, we design the coefficients for SNR $=10$ dB which can be regarded as relatively large SNR compared to the variance of interference. For this case, the optimal power allocation $\tilde{P}_a$ for both $\rho = 0.1$ and $\rho = 0.5$ are relatively small. Therefore, the performance improvement provided by the HDA scheme is larger than that provided by the digital DPC scheme for both cases.

In Fig.~\ref{fig:diff_slopes}, we plot the performance of the proposed schemes with different choices of $P_a$ for the same channel parameters with those in the previous figure for $\rho = 0.1$. We observe that for both schemes, if we compromise the optimality at the designed SNR, it is possible to get better slopes of distortion than that obtained by setting $P_a = \tilde{P}_a$. In other words, we can obtain a family of achievable distortion under SNR mismatch by choosing $P_a\in [0,P]$.

%%%%%%%%%%%%%%%%%%%%%%%%%%%%%%%%%%%%%%%%%%%%%%%%%%%%%%%%%%%%%%%
\section{JSCC for the Generalized Cognitive Radio Channel}\label{sec:JSCC_cograd}
An interesting application of the joint source-channel coding problem considered in this paper is in the transmission of analog sources over a cognitive radio channel. In this section, we will first formally state the problem, derive an outer bound on the achievable distortion region, and then propose a coding scheme based on the schemes given in Section~\ref{sec:proposed_scheme}.

\subsection{Problem Statement}
Recently, there has been a lot of interest in cognitive radio since it was proposed in \cite{Mitola} for flexible communication devices and higher spectral efficiency. In a conventional cognitive radio setup, the lower priority user  (usually referred to as the secondary user) listens to the wireless channel and transmits the signal only through the spectrum not used by the higher priority user (referred to as the primary user). In a generalized cognitive radio channel, simultaneous transmission over the same time and frequency is allowed. As shown in Fig.~\ref{fig:sys_mod_cogradio}, the problem can be modeled as an interference channel with direct channel gain $1$ and cross channels $h_1$ and $h_2$ representing the real-valued channel gains from user $1$ to user $2$ and vice versa, respectively. The average power constraints imposed on the outputs of user 1 and 2 are $P_1$ and $P_2$, respectively. Different from interference channels, in cognitive radio channels, we further assume that the secondary user knows $V_1$ non-causally. Here, we also assume that the channel coefficient $h_1$ is known by the secondary user. The received signals are given by
\begin{equation}
    \left(
      \begin{array}{c}
        Y_1 \\
        Y_2 \\
      \end{array}
    \right)
    =
    \left(
      \begin{array}{cc}
        1 & h_1 \\
        h_2 & 1 \\
      \end{array}
    \right)
    \left(
      \begin{array}{c}
        X_1 \\
        X_2 \\
      \end{array}
    \right)
    +
    \left(
      \begin{array}{c}
        Z_1 \\
        Z_2 \\
      \end{array}
    \right).
\end{equation}
where $Z_i\sim \mathcal{N}(0,1)$ for $i\in\{1,2\}$. The capacity region of this channel has been studied and is known for some special cases, e.g., the weak interference case \cite{WuVishArap} \cite{JovVis}, the very-strong interference case \cite{MaricYatesKramer}, and the primary-decode-cognitive case \cite{RiniTunDev}.

In this section, we consider the same generalized cognitive radio channel but our focus is on the case when both users have {\em analog} information $V_1$ and $V_2$, respectively. We are interested in the distortion region which describes how much distortion two users can achieve simultaneously. In particular, we consider the case when the two sources are correlated with a covariance matrix given by
\begin{equation}
    \Lambda_{V_1 V_2} = \left(
                 \begin{array}{cc}
                   \sigma_{V_1}^2 & \rho\sigma_{V_1}\sigma_{V_2} \\
                   \rho\sigma_{V_1}\sigma_{V_2} & \sigma_{V_2}^2 \\
                 \end{array}
               \right).
\end{equation}
The distortion measure is the MSE distortion measure defined in~\eqref{eqn:MSE}. An achievable distortion region can be obtained by first enforcing the primary user to use the uncoded scheme and using the proposed schemes given in section~\ref{sec:proposed_scheme} for the secondary user. In fact, since the primary user does not have any side-information, analog transmission is an optimal choice \cite{Goblick} \cite{GasRimVet} in terms of the distortion achieved at the primary receiver. Further notice that since we do not consider SNR mismatch here, it makes no difference which proposed scheme we use.

\subsection{Outer Bound}
In this subsection, we derive an outer bound on the distortion region for the generalized cognitive radio channel with $R_1 = I(X^n_1;Y^n_1)$ and $R_2 = I(X^n_2;Y^n_2|X^n_1)$. Then, for the primary user, we have
\begin{align}
    \frac{n}{2}\log\frac{\sigma_{V_1}^2}{D_1} &\overset{(a)}{\leq} I(V^n_1;\widehat{V}^n_1) \nonumber \\
    &\overset{(b)}{\leq} I(X^n_1;Y^n_1) \nonumber \\
    &=nR_1,
\end{align}
where (a) follows from rate distortion theory and (b) follows from the data processing inequality. Also, for the secondary user, we have
\begin{align}
    \frac{n}{2}\log\frac{\sigma_{V_2}(1-\rho^2)}{D_2} &\leq I(V^n_2;\widehat{V}^n_2|V^n_1) \nonumber \\
    &\overset{(a)}{=}I(V^n_2;\widehat{V}^n_2|V^n_1,X^n_1)\nonumber \\
    &\overset{(b)}{=}I(V^n_2;\widehat{V}^n_2|X^n_1)\nonumber \\
    &\overset{(c)}{\leq} I(X^n_2;Y^n_2|X^n_1)\nonumber \\
    &=nR_2,
\end{align}
where (a) is due to the fact that $X^n_1$ is a deterministic function of $V^n_1$, (b) follows from the Markov chain $V^n_1\ \leftrightarrow(X^n_1,X^n_2)\leftrightarrow(Y^n_1,Y^n_2)$, and (c) follows from the data processing inequality. Thus, we have
\begin{align}
    D_{ob1} &= \frac{\sigma_{V_1}^2}{R_1}, \label{eqn:ob1}\\
    D_{ob2} &= \frac{\sigma_{V_2}^2(1-\rho^2)}{R_2}, \label{eqn:ob2}
\end{align}
where $(R_1,R_2)$ must lie inside the capacity region of the generalized cognitive radio channel.

As mentioned earlier, the capacity region of this channel setup is only known for some special cases. Fortunately, for those cases whose capacity regions remain unknown, outer bounds on $(R_1,R_2)$ are available (see e.g. \cite{RiniTunDev} wherein the authors give an unified view of outer bounds for different cases) and therefore we can still obtain the outer bound given in \eqref{eqn:ob1} and \eqref{eqn:ob2}.

\subsection{Proposed Coding Scheme}
Let the primary user simply transmit the scaled version of the uncoded source $X_1 = \sqrt{P_1/\sigma_{V_1}^2}V_1$. Therefore, the bottom channel in Fig.~\ref{fig:sys_mod_cogradio} reduces to the situation we considered in the previous section with source $V = V_2$ and interference $S = h_1 X_1$. The covariance matrix becomes \eqref{eqn:CovSV} with
\begin{align}
    \sigma_V^2 &= \sigma_{V_2}^2, \\
    \sigma_S^2 &= h_1^2 P_1.
\end{align}
The secondary user then encodes its source to $X_2$ by the HDA scheme described previously in section \ref{subsec:HDA} with power $P_2 = P_h + P_a$ and coefficients according to \eqref{eqn:alp}. With these coefficients, the corresponding distortion $D_2$ is computed by \eqref{eqn:D_hda}. At the receiver 1, the received signal is
\begin{align}
    Y_1 &= X_1 + h_2 X_2 + Z_1 \nonumber \\
    &= \left( 1 + (1-\gamma)\sqrt{a}h_1 h_2 \right)X_1 + h_2X_h + h_2\sqrt{a}\gamma V_2 + Z_1.
\end{align}
Decoder 1 then forms a linear MMSE estimate from $Y_1$ given by $\widehat{V}_1 = \beta_1 Y_1$, where $\beta_1 = \mathbb{E}[V_1 Y_1]/\mathbb{E}[Y_1^2]$
and
\begin{align}
    \mathbb{E}[V_1 Y_1] &= \left( 1 + (1-\gamma)\sqrt{a}h_1 h_2 \right)\sqrt{P_1\sigma_{V_1}^2} + h_2 \sqrt{a}\gamma\rho\sigma_{V_1}\sigma_{V_2} \\
    \mathbb{E}[Y_1^2] &= \left( 1 + (1-\gamma)\sqrt{a}h_1 h_2 \right)^2 P_1 + a h_2^2\gamma^2\sigma_{V_2}^2 +  \nonumber \\
    &h_2^2P_h+2\sqrt{a}h_2\gamma\rho\sqrt{P_1\sigma_{V_2}^2}\left( 1 + (1-\gamma)\sqrt{a}h_1 h_2 \right) + N_1.
\end{align}
Therefore, the corresponding distortion is $D_1 = \sigma_{V_1}^2 - \beta_1 \mathbb{E}[V_1Y_1]$.

It can be verified that assigning $\gamma = 1$ may lead to a suboptimal $D_1$ in general. Thus, as we mentioned in Section~\ref{subsec:DigitalDPC}, one may want to assign a non-zero power to transmit $S$ in order to achieve a larger distortion region. We can then optimize the power allocation for particular performance criteria. For instance, if one desires achieving the minimum distortion for the secondary user, $\gamma$ should be set to be $1$. However, if the aim is to obtain the largest achievable distortion region, one should optimize over $P_a\in[0,P_1]$ and $\gamma\in[0,1]$.

\subsection{Discussions and Numerical Results}
Here, we give examples to compare the performance of the outer bound and the proposed coding scheme for two cases whose capacity region is known, namely the weak interference case and very-strong interference case. Also, similar to \cite{JovVis}, we also present the distortion for the secondary user under the coexistence conditions.

\emph{1. Weak interference case}: When the interference is weak, i.e., $|h_2|\leq 1$, the capacity region is given by \cite{WuVishArap} \cite{JovVis}
\begin{align}
    R_1&\leq \frac{1}{2}\log\left(1+\frac{P_1(1+h_2\rho_x\sqrt{P_2/P_1})^2}{1+(1-\rho_x^2)h_2^2P_2}\right) \nonumber \\
    R_2&\leq \frac{1}{2}\log\left(1+(1-\rho_x^2)P_2\right),
\end{align}
where $\rho_x\in[0,1]$. One can see that the capacity region of this case is a rectangle; therefore, increasing $R_2$ will not affect $R_1$. For this case, the outer bounds in \eqref{eqn:ob1} and \eqref{eqn:ob2} become
\begin{align}
    D_{ob1} &= \frac{\sigma_{V_1}^2}{1+\frac{P_1(1+h_2\rho_x\sqrt{P_2/P_1})^2}{1+(1-\rho_x^2)h_2^2P_2}}, \\
    D_{ob2} &= \frac{\sigma_{V_2}^2(1-\rho^2)}{1+(1-\rho_x^2)P_2}.
\end{align}

One example of the distortion region for this case is shown in Fig.~\ref{fig:weak} in which we plot the outer bound and the boundary of the distortion region achieved by the proposed coding scheme. The parameters are set to be $\sigma_{V_1}^2 = \sigma_{V_2}^2 = 1$, $h_1 = h_2 = 0.5$, and the power constraints are $P_1 = P_2 = 1$. In this figure,  One can observe that when $\rho=0$, the outer bound is tight and the proposed coding scheme is optimal. However, the inner and outer bound do not coincide for other $\rho$s and one can see that the gap increases as $\rho$ increases.

\emph{2. Very-strong interference case}: The channel is said to be in the very-strong interference regime if the following conditions are satisfied,
\begin{equation}\label{eqn:very_strong1}
    |h_2|\geq 1,
\end{equation}
\begin{equation}
    |h_1\sqrt{P_1/P_2}+1|\geq|\sqrt{P_1/P_2}+h_2|,
\end{equation}
\begin{equation}\label{eqn:very_strong3}
    |h_1\sqrt{P_1/P_2}-1|\geq|\sqrt{P_1/P_2}-h_2|.
\end{equation}
The capacity region of this case is the union of $(R_1,R_2)$ satisfying \cite{MaricYatesKramer}
\begin{align}
    R_2 &\leq \frac{1}{2}\log\left(1+(1-\rho_x^2)P_2\right) \nonumber \\
    R_1+R_2 &\leq \frac{1}{2}\log\left(1 + P_1 + h_2^2P_2+2\rho_x h_2\sqrt{P_1P_2})^2 \right),
\end{align}
where $\rho_x\in[0,1]$. For this case, different choices of $R_2$ may lead to different upper bounds for $R_1$. Thus, the outer bound can be obtained by collecting all the Pareto minimal points of $(D_1,D_2)$ among all choices of $(R_1,R_2)$ and $\rho_x$.

In Fig.~\ref{fig:strong}, the outer bound and the boundary of the distortion region achieved by the proposed scheme are plotted. All the parameters are set to be the same as those in the previous figure except for $h_1 = h_2 = 1.5$ now. It is easy to see that \eqref{eqn:very_strong1}-\eqref{eqn:very_strong3} are satisfied under these parameters. One can see that for this case the inner and outer bound do not coincide even for $\rho=0$ case. This may be due to the fact that in the proposed coding scheme, the primary decoder treats the signal from the secondary user as extra noise. This violates the insight of the very-strong interference regime that one should first decode interfering signal and then cancel it out since the interference is ``very-strong" and is regarded as easier to decode. However, for the proposed scheme, the primary decoder is not able to obtain an improvement from this decoding strategy. This is because the digital part (or the HDA part, depends on which scheme is used) of the interfering signal is a function of $V_1$ and the bin index (or $U$). Therefore, decoding the bin index (or $U$) only is not enough to reconstruct $X_d$ (or $X_h$).

On the other hand, if one simply ignores the correlation and uses an optimal separate source-channel code at the secondary user, this coding scheme is guaranteed to achieve the outer bound for $\rho=0$ but this scheme is unable to adapt with $\rho$, i.e., the performance is fixed for all $\rho$s. Therefore, when $\rho$ is large, one may obtain a lower distortion by using the proposed scheme although it fails to achieve the outer bound for any $\rho$. One example is given in Fig.~\ref{fig:strong} that when $\rho=0.5$, the distortion region achieved by the proposed scheme is larger than that achieved by an optimal separate coding scheme (whose performance is the same as the outer bound for $\rho = 0$). It is interesting to build a coding scheme that achieves the outer bound for $\rho=0$ and is capable of adapting with $\rho$ for the very-strong interference case; however, this is beyond the scope of this paper.

\emph{3. Coexistence Conditions}:
In \cite{JovVis}, the coexistence conditions are introduced to understand the system-wise benefits of cognitive radio. The authors study the largest rate that the cognitive radio can achieve under these coexistence conditions described as follows.

\emph{1}. the presence of cognitive radio should not create rate degradation for the primary user, and

\emph{2}. the primary user does not need to use a more sophisticated decoder than it would use in the absence of the cognitive radio. i.e, a single-user decoder is enough.

Similar to this idea, we study the distortion of the secondary user under the following conditions

\emph{1}. the presence of cognitive radio should not create distortion increment for the primary user, and

\emph{2}. the primary user uses a single-user decoder.

We present the outer bound and the signal-to-distortion ratio for the secondary user obtained by the proposed scheme under coexistence conditions. Here the outer bound is given by
\begin{equation}
    D_{ob2,coexist} = \underset{D_{ob1}\leq \frac{\sigma_{V_1}^2}{1+P_1}}{\inf} D_{ob2},
\end{equation}
where $D_{ob1}$ and $D_{ob2}$ are given in \eqref{eqn:ob1} and \eqref{eqn:ob2}, respectively, and $R_1$ and $R_2$ therein can be further bounded by the capacity region or upper bounds on the capacity region as mentioned. Note that when taking the infimum, we simply constrain the distortion of the primary user to be at most the one achieved when there is no interference at all and ignore the second coexistence condition. i.e., this outer bound allows the primary decoder to be any possible decoder, not necessary a single-user decoder.

In Fig.~\ref{fig:coexist_weak} and Fig.~\ref{fig:coexist_strong}, the achievable distortion for the secondary user is plotted for the same set of parameters as in Fig.~\ref{fig:weak} and Fig.~\ref{fig:strong}, respectively. As shown in these figures, the proposed scheme is able to increase the secondary user's signal-to-distortion ratio without degrading the performance of the primary user. Moreover, one can observe that at $\rho=0$ the proposed coding is optimal for the weak interference case but not for the very-strong interference case. This may be due to the fact that in the proposed coding scheme the interfering signal is not fully decoded. This may also be the consequence of ignoring the second condition when deriving the outer bound. Another interesting observation is that in Fig.~\ref{fig:coexist_strong}, the signal-to-distortion ratio increases more rapidly than that in Fig.~\ref{fig:coexist_weak}. This is because in the very-strong interference case, the channel would amplify the secondary user's signal much more than that in the weak interference case. So the secondary user could use less power to boost the primary signal such that the coexistence conditions are satisfied and then use the remaining power to decrease its own distortion.

\section{Conclusions}\label{sec:conclusions}
In this paper, we have discussed the joint source-channel coding problem with interference known at the transmitter. In particular, we considered the case that the source and the interference are correlated with each other. We proposed a digital DPC scheme and a HDA scheme and showed that both two schemes can adapt with $\rho$. The performance of these two schemes under SNR mismatch are also discussed. Different from typical separation-based schemes which are not able to take advantage of a better channel SNR and suffer from abrupt degradation when the channel deteriorates, both the proposed schemes can benefit from a better side-information acquired at the decoder and also provide a graceful degradation and improvement under SNR mismatch. However, there is a difference between the performance of the two proposed schemes when $\text{SNR}_a >\text{SNR}_d$ and which scheme is better depends on the designed SNR and $\rho$.

These two schemes are then applied to the generalized cognitive radio channel for deriving an achievable distortion region. Outer bounds on distortion region for this channel are also provided. To the best of our knowledge, this is the first joint source-channel coding scheme that has been proposed for the generalized cognitive radio channel. Numerical results suggest that, in the weak interference regime, the gap between the inner and outer bound is reasonably small for small and medium $\rho$ and increases as $\rho$ increases. Moreover, in the very-strong interference regime, there exist $\rho$s such that the proposed joint source-channel coding scheme outperforms optimal separate coding scheme. The system-wise benefits of cognitive radio in terms of distortion are also studied via imposing the coexistence conditions.

\appendices
\section{Equivalence of \eqref{eqn:D_sep} and \eqref{eqn:D_hda}}\label{app:same}
In this appendix, we verify that with the knowledge of actual channel SNR, two proposed schemes perform exactly the same. For fixed $\gamma$ and $P_h = P - P_a$, the second term in \eqref{eqn:D_hda} becomes
\begin{equation}\label{eqn:GLG}
    \Gamma^T \Lambda_{UY}^{-1}\Gamma = \frac{\mathbb{E}[VU]^2\mathbb{E}[Y^2]-2\mathbb{E}[VU]\mathbb{E}[VY]\mathbb{E}[UY]+\mathbb{E}[VY]^2\mathbb{E}[U^2]}{\mathbb{E}[U^2]\mathbb{E}[Y^2]-\mathbb{E}[UY]^2},
\end{equation}
where
\begin{equation}
    \mathbb{E}[VU] = \alpha\mathbb{E}[S'V]+\kappa\sigma_V^2,
\end{equation}
\begin{equation}
    \mathbb{E}[VY] = \mathbb{E}[S'V],
\end{equation}
\begin{equation}
    \mathbb{E}[U^2] = P_h + \alpha^2 \mathbb{E}[S'^2] + \kappa^2\sigma_V^2 + 2\alpha\kappa\mathbb{E}[S'V],
\end{equation}
\begin{equation}
    \mathbb{E}[Y^2] = P_h + \mathbb{E}[S'^2] + N,
\end{equation}
\begin{equation}
    \mathbb{E}[UY] = P_h + \alpha\mathbb{E}[S'^2] + \kappa\mathbb{E}[S'V],
\end{equation}
with $\alpha$ and $\kappa^2$ determined by \eqref{eqn:alp} and
\begin{equation}
    \mathbb{E}[S'^2] = a\gamma^2\sigma_V^2 + [1 + \sqrt{a}(1-\gamma)]^2\sigma_S^2 + 2\sqrt{a}\gamma[1 + \sqrt{a}(1-\gamma)]\rho\sigma_V\sigma_S,
\end{equation}
\begin{equation}
    \mathbb{E}[S'V] = \sqrt{a}\gamma\sigma_V^2 + [1 + \sqrt{a}(1-\gamma)]\rho\sigma_V\sigma_S.
\end{equation}

After some algebra, we can rewrite the numerator and denominator in \eqref{eqn:GLG} as, respectively,
\begin{equation}\label{eqn:numerator}
    \frac{P_h\left\{\mathbb{E}[S'V]^2ND^* + \sigma_V^2P_h(\sigma_V^2\mathbb{E}[Y^2]-\mathbb{E}[S'V]^2)\right\}}{(P_h+N)D^*},
\end{equation}
and
\begin{equation}\label{eqn:denominator}
    \frac{P_h\left\{ (ND^*+P_h\sigma_V^2)\mathbb{E}[Y^2] - P_h\mathbb{E}[S'V]^2 \right\}}{(P_h+N)D^*}.
\end{equation}
Thus, we can rewrite \eqref{eqn:D_hda} as
\begin{align}
    D_{hda} &= \sigma_V^2 - \Gamma^T \Lambda_{UY}^{-1}\Gamma = \sigma_V^2 - \frac{\eqref{eqn:numerator}}{\eqref{eqn:denominator}} \nonumber \\
    &= D^{*}\frac{N(\sigma_V^2\mathbb{E}[Y^2]-\mathbb{E}[S'V]^2)}{(ND^*+P_h\sigma_V^2)\mathbb{E}[Y^2]-P_h\mathbb{E}[S'V]^2} \nonumber \\
    &\overset{(a)}{=}\frac{ND^*(\sigma_V^2\mathbb{E}[Y^2]-\mathbb{E}[S'V]^2)}{(P_h+N)(\sigma_V^2\mathbb{E}[Y^2]-\mathbb{E}[S'V]^2)} \nonumber \\
    &= \frac{D^*}{1 + \frac{P_h}{N}},
\end{align}
where (a) follows from that $D^* = \sigma_V^2 - \mathbb{E}[VY]^2/\mathbb{E}[Y^2]$ and $\mathbb{E}[VY] = \mathbb{E}[S'V]$. This completes the proof.

\section{Digital Wyner-Ziv Scheme}\label{app:DigitalWZ}
In this appendix, we summarize the digital Wyner-Ziv scheme for lossy source coding with side-information $V'$ ($V = V' + W$ with $W\sim\mathcal{N}(0,D^*)$) at receiver. Similar to the previous sections, we omit all the $\varepsilon$ and/or $\delta$ intentionally for the sake of convenience and to maintain clarity.

Suppose the side-information is available at both sides, the least required rate $R_{WZ}$ for achieving a desired distortion $D$ is \cite{Makesh}
\begin{equation}\label{eqn:WZ}
    R_{WZ} = \frac{1}{2}\log\frac{D^*}{D}.
\end{equation}
Let us set this rate to be arbitrarily close to the rate given in \eqref{eqn:channel_rate}, the rate that the channel can support with arbitrarily small error probability. The best possible distortion one can achieve for this setup is then given as
\begin{equation}
    D = \frac{D^*}{1 + \frac{P-P_a}{N}}.
\end{equation}
This distortion can be achieved as follows \cite{Makesh},

    \emph{1}. Let $T$ be the auxiliary random variable given by
    \begin{equation}\label{eqn:auxiliaryT}
        T = \alpha_{sep}V+B,
    \end{equation}
    where
    \begin{equation}\label{eqn:alp_sep}
        \alpha_{sep} = \sqrt{\frac{D^*-D}{D^*}}
    \end{equation}
    and $B\sim \mathcal{N}(0,D)$. Generate a length $n$ i.i.d. Gaussian codebook $\mathcal{T}$ of size $2^{nI(T;V)}$ and randomly assign the codewords into $2^{nR}$ bins with $R$ chosen from \eqref{eqn:channel_rate}. For each source realization $\mathbf{v}$, find a codeword $\mathbf{t}\in\mathcal{T}$ such that $(\mathbf{v},\mathbf{t})$ is jointly typical. If none or more than one are found, an encoding failure is declared.

    \emph{2}. For each chosen codeword, the encoder transmit the bin index of this codeword by the DPC with rate given in \eqref{eqn:channel_rate}.

    \emph{3}. The decoder first decodes the bin index (the decodability is guaranteed by the rate we chose) and then looks for a codeword $\hat{\mathbf{t}}$ in this bin such that $(\hat{\mathbf{t}},\mathbf{v}')$ is jointly typical. If this is not found, a dummy codeword is selected. Note that as $n\rightarrow \infty$, the probability that $\hat{\mathbf{t}}\neq \mathbf{t}$ vanishes. Therefore, we can assume that $\hat{\mathbf{t}}=\mathbf{t}$ from now on.

    \emph{4}. Finally, the decoder forms the MMSE from $\mathbf{t}$ and $\mathbf{v}'$ as $\hat{\mathbf{v}} = \mathbf{v}' + \hat{\mathbf{w}}$ with
    \begin{equation}
        \hat{\mathbf{w}} = \frac{\alpha_{sep}D^*}{\alpha_{sep}^2 D^* + D} (\mathbf{t} - \alpha_{sep}\mathbf{v}').
    \end{equation}
It can be verified that for the choice of $\alpha$ the required rate is equal to \eqref{eqn:WZ} and the corresponding distortion is
\begin{align}
    &\mathbb{E}[(V-\widehat{V})^2]=\mathbb{E}[(W-\widehat{W})^2]\nonumber \\
    &= D^* \left( 1 - \frac{\alpha_{sep}^2 D^*}{\alpha_{sep}^2 D^* + D}\right)= D.
\end{align}

\section{Wyner-Ziv with Mismatched Side-Information}\label{app:WZmismatch}
In this appendix, we calculate the expected distortion of the digital Wyner-Ziv scheme in the presence of side-information mismatch. Specifically, we consider the Wyner-Ziv problem with an i.i.d. Gaussian source and the MSE distortion measure. Let us assume that the best achievable distortion in the absence of side-information mismatch to be $D$. The encoder believes that the side-information is $V'$, and $V = V'+ W$ with $W \sim N(0,D^*)$. However, the side-information turns out to be $V'_a$ and has the relation $V = V'_a + W_a$ with $W_a\sim N(0,D_a^*)$. Under the same rate, we want to calculate the actual distortion $D_a$ suffered by the decoder.

Since the encoder has been fixed to deal with the side-information, $V'$, at decoder, the auxiliary random variable is as in \eqref{eqn:auxiliaryT} with the coefficient given in \eqref{eqn:alp_sep}. Since the decoder knows the actual side-information, $V'_a$, perfectly, it only has to estimate $W_a$. By the orthogonality principle, the MMSE estimate $\widehat{W}_a$ can be obtained as
\begin{equation}
    \widehat{W}_a=\frac{\alpha_{sep} D_a^*}{\alpha_{sep}^2 D_a^*+D} (T - \alpha_{sep} V'_a)
\end{equation}
Therefore, the estimate of the source is $\widehat{V} = V'_a+\widehat{W}_a$.
The corresponding distortion is given as
\begin{align}
D_a &= \mathbb{E}[(V-\widehat{V})^2]=\mathbb{E}[(W_a-\widehat{W}_a)^2]\nonumber \\
    &=\frac{D^*D^*_a}{D^* D^*_a + (D^*-D^*_a) D} D
\end{align}

Here, we give an example in Fig.~\ref{fig:SI_mismatch} to see the performance improvement through having the access of a better side-information. In this figure, we plot the $-10\log_{10}D_a$ as $-10\log_{10} D^*_a$ increases, i.e., as the actual side-information improves. The outer bound is obtained by assuming the transmitter always knows the distribution of actual side-information at decoder and the distortion of the HDA scheme is computed through derivations in section \ref{subsec:HDA}. The parameters are set to be $P = N = 1$ and $D^* = 0.1$. One can observe in the figure that both the schemes benefit from a better side-information at decoder. Moreover, it can be seen that these two schemes provide the same performance under side-information mismatch.

\section{Discussions for SNR Mismatch Cases}\label{app:DiscussMismatch}
As discussed previously, both the digital DPC scheme and the HDA scheme benefit from a better SNR. Here, we wish to analyze and compare the performance for these two schemes under SNR mismatch. Since the digital DPC scheme makes estimate from $T$ (see Appendix \ref{app:DigitalWZ}) and $V'$ (which is a function of $Y$) and the HDA scheme makes estimate from $U$ and $Y$, it suffices to compare $I(V;T,Y)$ with $I(V;U,Y)$. By the chain rule of mutual information, we have
\begin{equation}
    I(V;T,Y) = I(V;Y) + I(V;T|Y),
\end{equation}
and
\begin{equation}
    I(V;U,Y) = I(V;Y) + I(V;U|Y).
\end{equation}
Thus, we only have to compare $I(V;T|Y)$ to $I(V;U|Y)$. Let us consider $\rho =0$ case for example,
\begin{align}\label{eqn:mutual_T}
    I(V;T|Y) &= h(T|Y) - h(T|V,Y) \nonumber \\
    &= h(\alpha_{sep}V + B |Y) - h(\alpha_{sep}V + B |V,Y) \nonumber \\
    &= h(\alpha_{sep} V - \alpha_{sep}\beta_a Y + B |Y) - h(\alpha_{sep} B |V,Y) \nonumber \\
    &= h(\alpha_{sep} W_a + B |Y) - h(B) \nonumber \\
    &\overset{(a)}{=} h(\alpha_{sep} W_a + B) - h(B) \nonumber \\
    &= \frac{1}{2}\log\frac{\alpha_{sep}^2 D_a^*+D}{D},
\end{align}
where $\alpha_{sep}$ and $W_a$ are defined in Appendix \ref{app:WZmismatch} and (a) follows from the orthogonality principle.
\begin{align}\label{eqn:mutual_U}
    I(V;U|Y) &= h(U|Y) - h(U|V,Y) \nonumber \\
    &= h(U|Y) - h(X_h + \alpha S' + \kappa V|V,Y) \nonumber \\
    &= h(U|Y) - h\left( (1-\alpha)X_h -\alpha Z_a|V,Y \right) \nonumber \\
    &\overset{(a)}{\geq} h(U|Y) - h\left( (1-\alpha)X_h -\alpha Z_a \right) \nonumber \\
    &= \frac{1}{2} \log\frac{\mathbb{E}[U^2]-\mathbb{E}[UY]^2/\mathbb{E}[Y^2]}{(1-\alpha)^2P_h+\alpha^2 N_a}.
\end{align}
where (a) follows from that conditioning reduces entropy and the equality occurs if there is no SNR mismatch.

Two examples are given in Fig.~\ref{fig:mutual_compare} to compare these two quantities with and without SNR mismatch for a small and a large designed SNR, respectively. One can observe that without SNR mismatch, these two quantities coincide with each other for all choices of $P_a$. This implies the result in section \ref{sec:proposed_scheme} that without mismatch the digital DPC scheme and the HDA scheme provide exactly the same distortion. However, with SNR mismatch, we can observe that which quantity is larger really depends on $P_a$ for the small designed SNR case. On the other hand for designed SNR = 10 dB case, we have $I(V;U|Y)>I(V;T|Y)$ for a wide range of $P_a$ (except for some $P_a$ close to 1). This explains the results in section \ref{sec:SNR_mismatch} that, for large designed SNRs, the HDA scheme has better results than the digital DPC scheme does while for small designed SNRs we cannot make this conclusion easily.

\begin{newpage}
\begin{figure}
    \centering
    \includegraphics[width=3in]{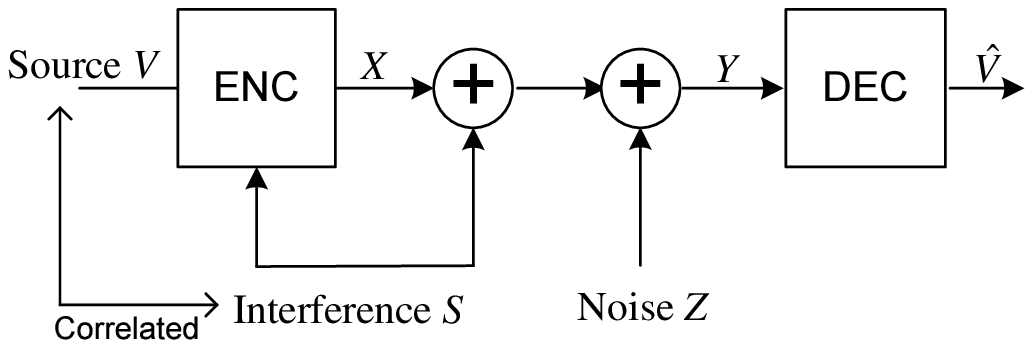}
    \caption{Joint source-channel coding with interference known at transmitter.}
    \label{fig:sys_mod_1}
\end{figure}

\begin{figure}
    \centering
    \includegraphics[width=3in]{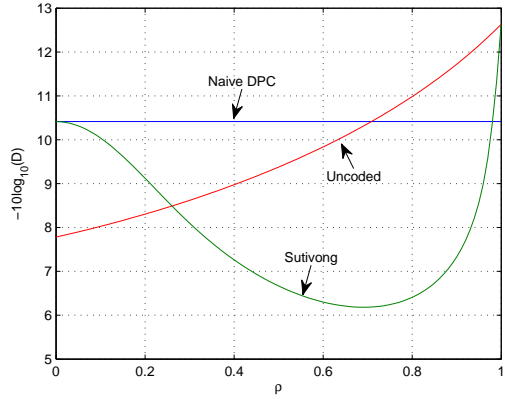}
    \caption{Distortions for the naive DPC, the uncoded scheme, and the extension of scheme in \cite{Sutivong}.}
    \label{fig:Sutivong}
\end{figure}

\begin{figure}
    \centering
    \includegraphics[width=2.7in]{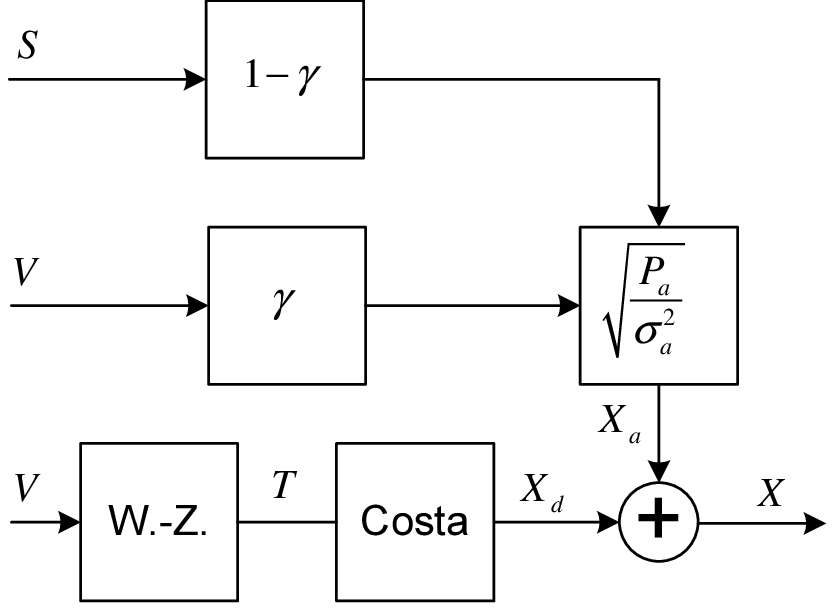}
    \caption{Digital DPC scheme.}
    \label{fig:sep_scheme}
\end{figure}

\begin{figure}
    \centering
    \includegraphics[width=3.5in]{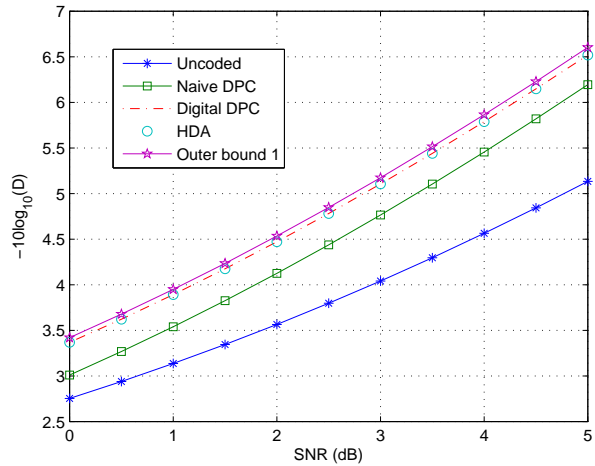}
    \caption{$\frac{P}{N}$ vs $D$, $\rho = 0.3.$}
    \label{fig:SNR_vs_D}
\end{figure}

\begin{figure}
    \centering
    \includegraphics[width=3.5in]{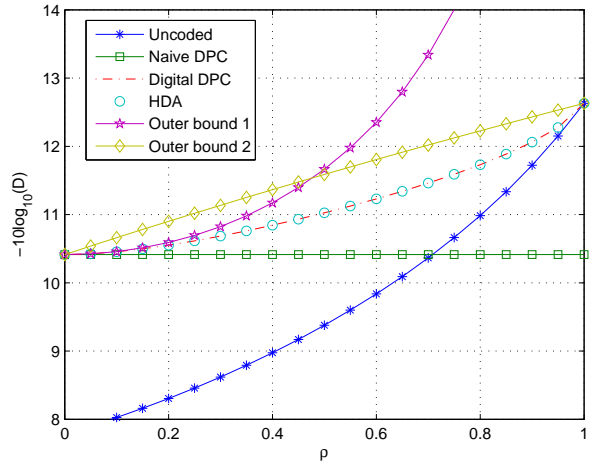}
    \caption{$\rho$ vs $D$ with $\sigma_V = \sigma_S = 1$ and $\frac{P}{N} = 10$.}
    \label{fig:rho_vs_D}
\end{figure}

\begin{figure}
    \centering
    \includegraphics[width=3.5in]{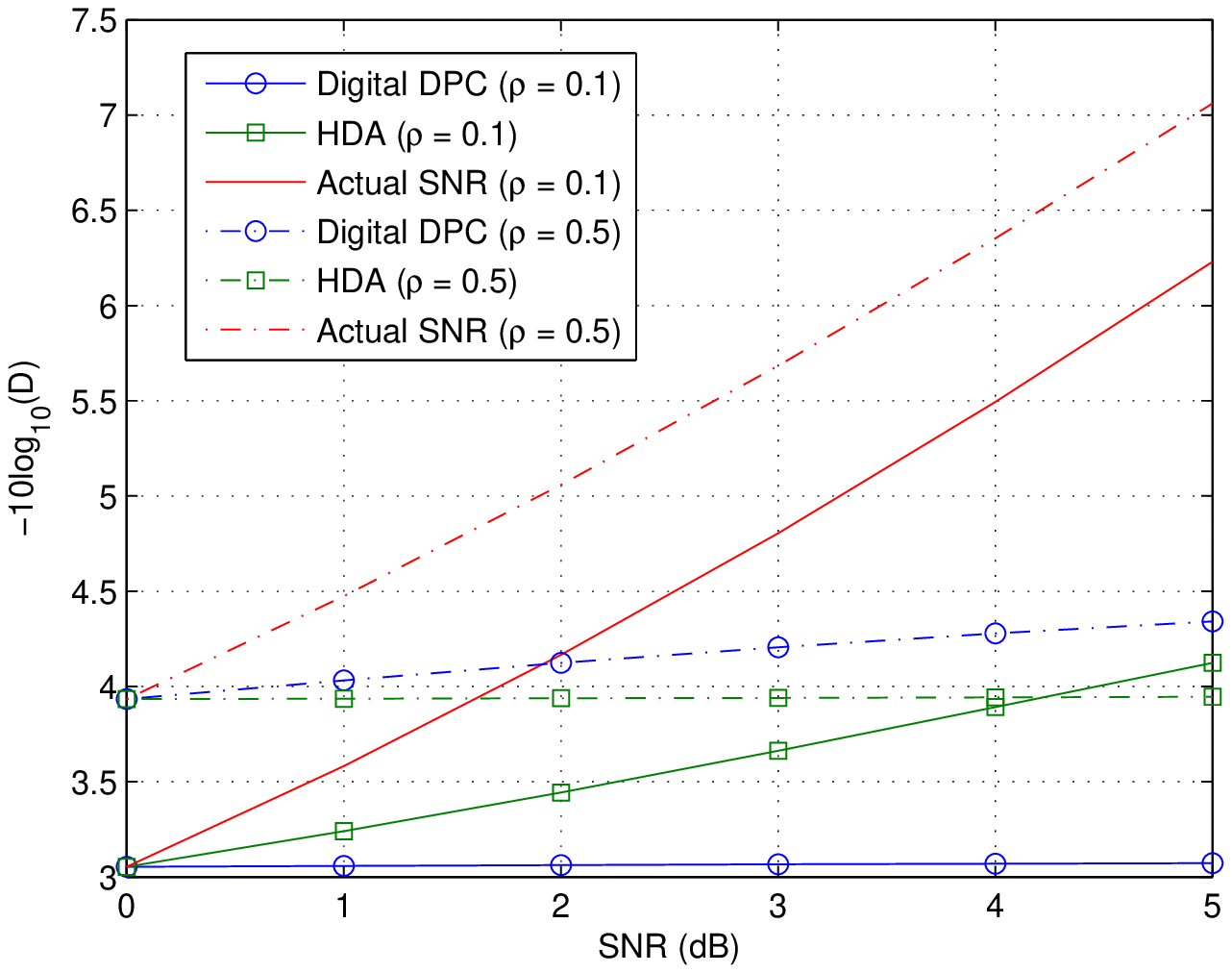}
    \caption{SNR mismatch case for SNR = 0dB.}
    \label{fig:JSCC_mismatch_small}
\end{figure}

\begin{figure}
    \centering
    \includegraphics[width=3.5in]{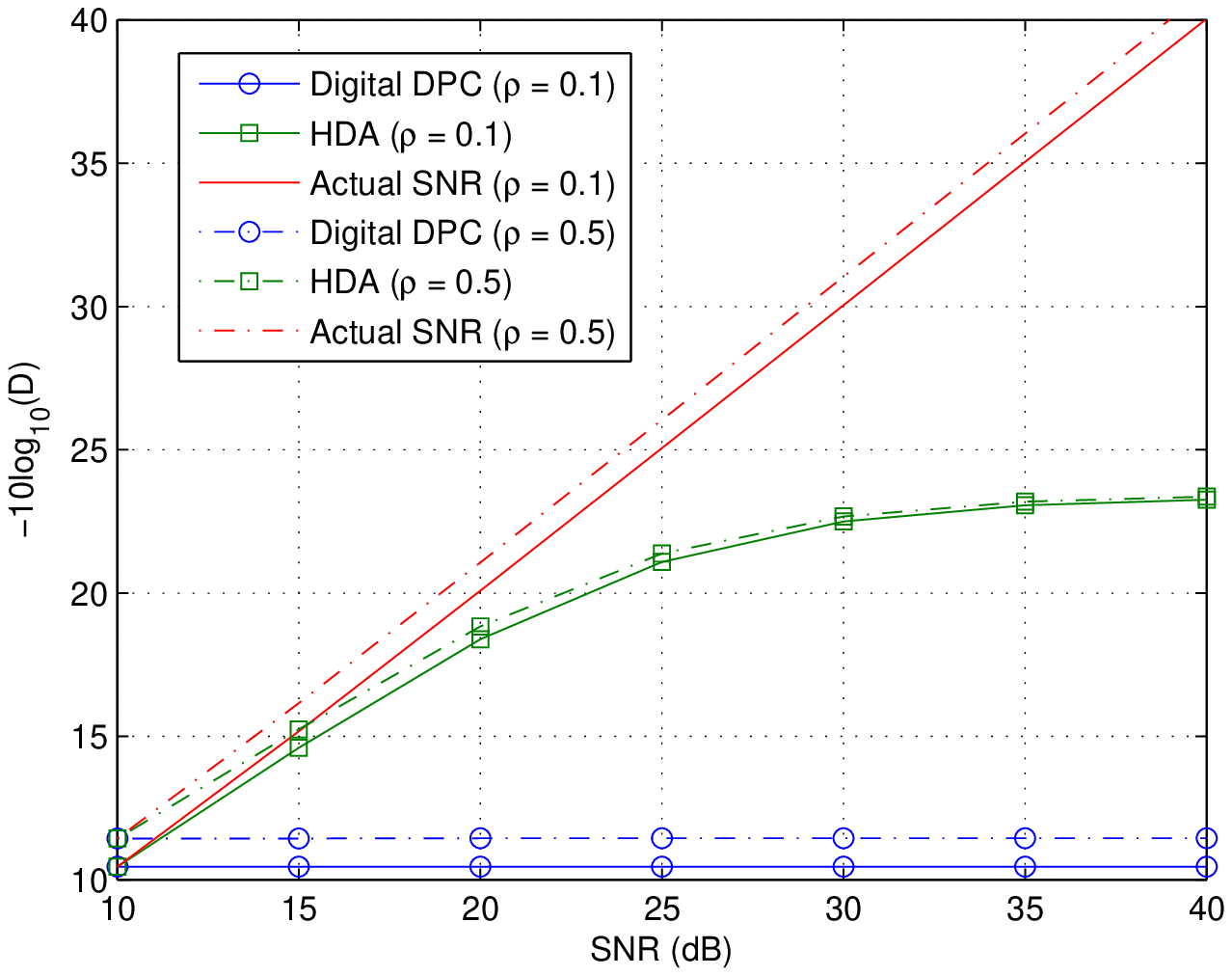}
    \caption{SNR mismatch case for SNR = 10dB.}
    \label{fig:JSCC_mismatch_large}
\end{figure}

\begin{figure}
    \centering
    \includegraphics[width=3.5in]{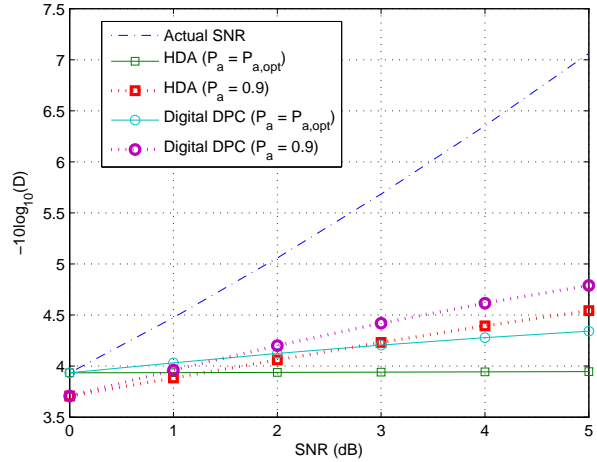}
    \caption{Proposed schemes with different choices of $P_a$.}
    \label{fig:diff_slopes}
\end{figure}

\begin{figure}
    \centering
    \includegraphics[width=3in]{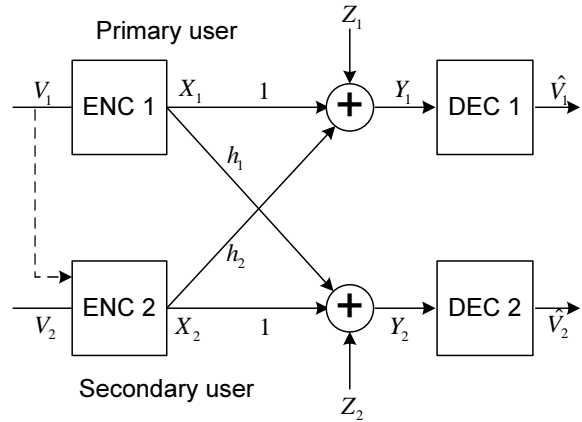}
    \caption{System model for a cognitive radio channel.}
    \label{fig:sys_mod_cogradio}
\end{figure}

\begin{figure}
    \centering
    \includegraphics[width=3.5in]{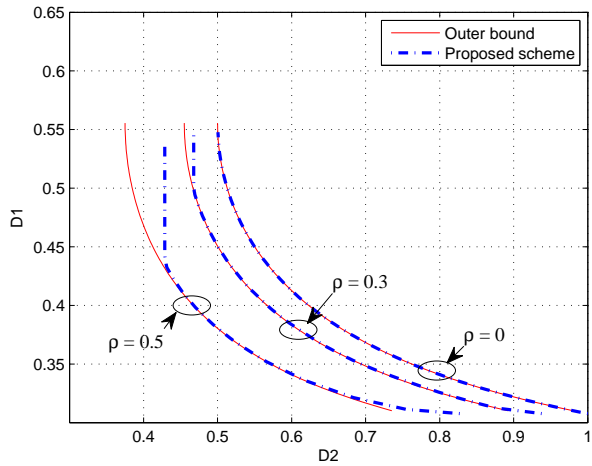}
    \caption{Distortion region for the weak interference case with $P_1=P_2=1,  \sigma^2_{V_1}=\sigma^2_{V_2}=1$, and $h_1=h_2=0.5$.}
    \label{fig:weak}
\end{figure}

\begin{figure}
    \centering
    \includegraphics[width=3.5in]{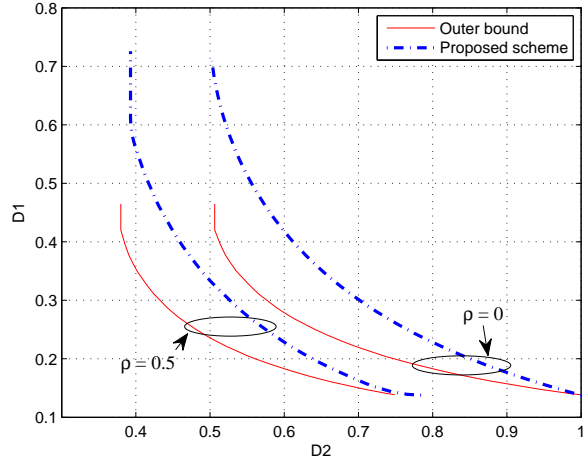}
    \caption{Distortion region for the very-strong interference case with $P_1=P_2=1,  \sigma^2_{V_1}=\sigma^2_{V_2}=1$, and $h_1=h_2=1.5$.}
    \label{fig:strong}
\end{figure}

\begin{figure}
    \centering
    \includegraphics[width=3.5in]{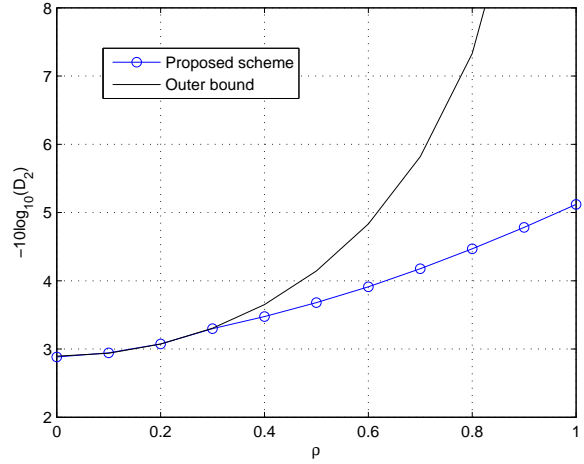}
    \caption{Distortion region for the weak interference case under coexistence conditions, $h_1=h_2=0.5$.}
    \label{fig:coexist_weak}
\end{figure}

\begin{figure}
    \centering
    \includegraphics[width=3.5in]{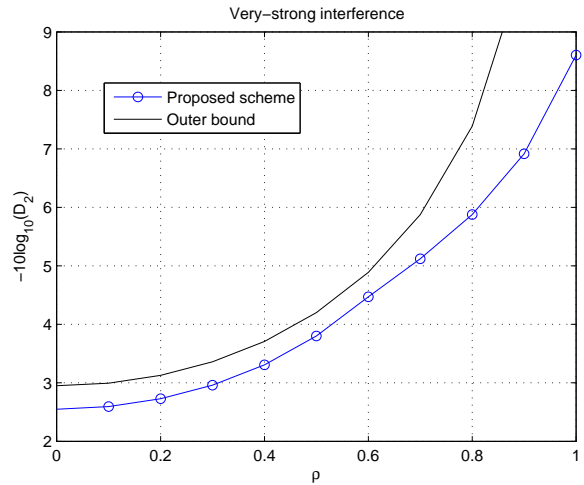}
    \caption{Distortion region for the very-strong interference case under coexistence conditions, $h_1=h_2=1.5$.}
    \label{fig:coexist_strong}
\end{figure}

\begin{figure}
    \centering
    \includegraphics[width=3.5in]{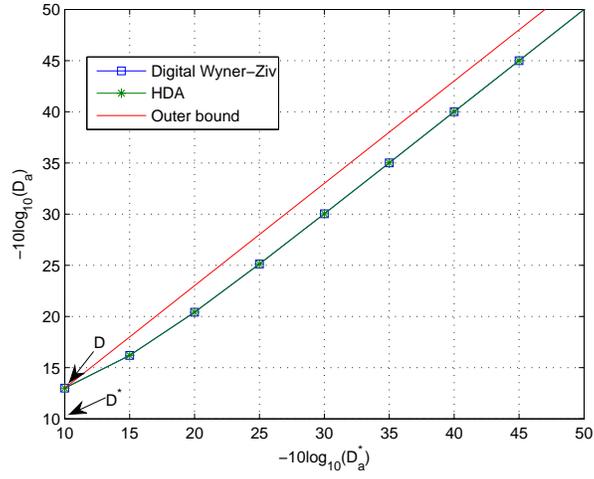}
    \caption{Wyner-Ziv problem with side-information mismatch.}
    \label{fig:SI_mismatch}
\end{figure}

\begin{figure}
    \centering
    \includegraphics[width=3.5in]{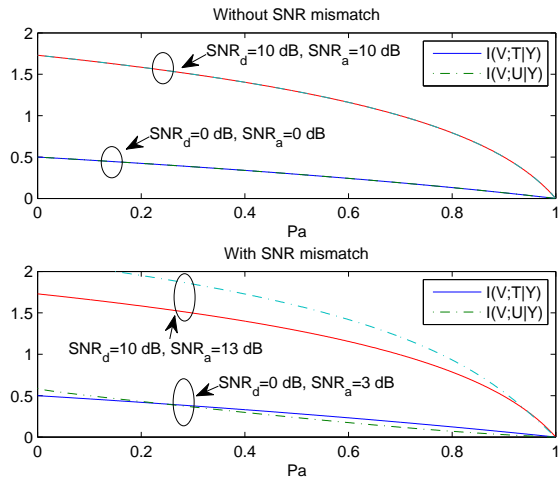}
    \caption{Comparisons of \eqref{eqn:mutual_T} and \eqref{eqn:mutual_U}.}
    \label{fig:mutual_compare}
\end{figure}

\end{newpage}

\end{document}